# Non-distributable key sharing for improving the security in IoT networks

**Mario Mastriani** [1,*]

[1] Knight Foundation School of Computing and Information Sciences, Florida International University, 11200 S.W. 8th Street, Miami, FL 33199, USA; mmastria@fiu.edu

[*] Correspondence: mmastria@fiu.edu

**Abstract:** Quantum key distribution (QKD) constitutes the most widespread family of information preservation techniques in the context of Quantum Cryptography. However, these techniques must deal with a series of technological challenges that prevent their efficient implementation, in space, as well as, in exclusively terrestrial configurations. Moreover, the current smallsat constellations of Low Earth Orbit (LEO with an altitude of approx. 500 km) added to other satellites in Medium Earth Orbit (MEO), and geostationary orbits (GEO), plus a large amount of space debris present around the planet, have become a serious obstacle for Astronomy. In this work, a classic alternative to QKD is presented, also based on a symmetric key, but with low cost, and high efficiency, which dispenses with all the implementation problems present in the QKD protocols and does not require the use of satellites, and which we will call non-distributable key sharing (NDKS). Due to its low cost, simplicity of implementation, and high efficiency, NDKS is presented as an ideal solution to the problem of cybersecurity in IoT, in general, and fog clouds, in particular.

## 1. Introduction

Internet of Things (IoT) is an emerging technology of technical, social, and economic importance [1]. Consumer products, durable goods, cars and trucks, industrial and utility components, sensors, and other everyday objects are now being combined with Internet connectivity and powerful data analytics capabilities, which promise to transform how everyday tasks such as work, life, and play are carried out [2]. The projections of the impact of the IoT on the Internet and the economy are remarkable. The best estimates anticipate that in the year 2025 there will be up to one hundred billion devices connected to the IoT and that its impact will be US$ 11 $10^{12}$. However, IoT also poses significant challenges that could make it difficult to realize its potential benefits [3]. News about attacks on Internet-connected devices, fears of surveillance, and privacy concerns have already captured the public's attention [4]. The technical challenges are still there, but new policy, legal and development challenges are also emerging. Definitively, its weak point is the security of the information and the preservation of the integrity of the data that travels through the network.

On the other hand, fog computing is the name of a cloud technology whereby the data generated by the devices is not uploaded directly to the cloud but is first prepared in smaller decentralized data centers [5]. The concept encompasses a network that extends from its borders, which is where the endpoints generate the data, to the central destination of the data in the public cloud or a private data center (private cloud) [6]. With the idea of a decentralized IT infrastructure, cloud computing would bring data processing closer to ground level. This is done with the so-called fog nodes, processing nodes before the cloud that acts as a mediator between the cloud and the different IoT devices [7]. The objective of fog computing also called fogging, is to shorten the communication paths between the cloud and the devices and reduce the data flow in external networks [8]. The nodes would thus fulfill an intermediate layer role in the network in which it is decided which data is processed locally and which is sent to the cloud or to a data center to be analyzed or processed.

As we have mentioned before, the problem that appears as the main stumbling block for the implementation of IoT techniques in a safe way, with or without the intervention of fog computing, is precisely the security of the information and the preservation of the integrity of the data. Therefore, given that all the current CyberSecurity tools used in IT have been widely violated, new cryptographic techniques based on principles of quantum mechanics [9] and grouped under the designation of Quantum Cryptography [10] appear to be a superior option. These tools are part of the arsenal of resources of Quantum Communications [11], whose maximum exponent is the future network of networks, i.e., the quantum Internet [12-17].

Precisely, the fundamental difference between the current Internet and the future Quantum Internet will lie in the security of information, which will be preserved in the latter through different families of techniques, among which the best known and applied is undoubtedly quantum key distribution (QKD) [10]. However, both in its purely terrestrial implementation [18, 19] as well as in space-to-ground or ground-to-space implementations [20-23], all QKD families have serious implementation problems.

These issues will be addressed in detail in the next section.

In summary, if we cannot do without QKD as far as information security is concerned, and yet its implementation problems persist in all its versions, then the only solution to ensure data integrity in IoT and the future quantum Internet is to resort to a technique that does not have the aforementioned problems. This is the central axis of this work by proposing a non-distributable key sharing (NDKS) technique to solve such an important problem.

The remainder of this paper is structured as follows. In Section 2, we describe the issues present in all QKD implementations. In Section 3, we discuss the non-distributable key sharing technique in detail. In Section 4, we provide a discussion and outline the limitations of our work. Finally, we conclude the paper by providing future research directions in Section 5.

## 2. Issues with QKD implementations

A typical QKD configuration requires two channels for the distribution of the key and an extra channel (public) for the transmission of the encrypted message. Figure 1 shows a generic QKD protocol, which consists of a symmetric-key system.

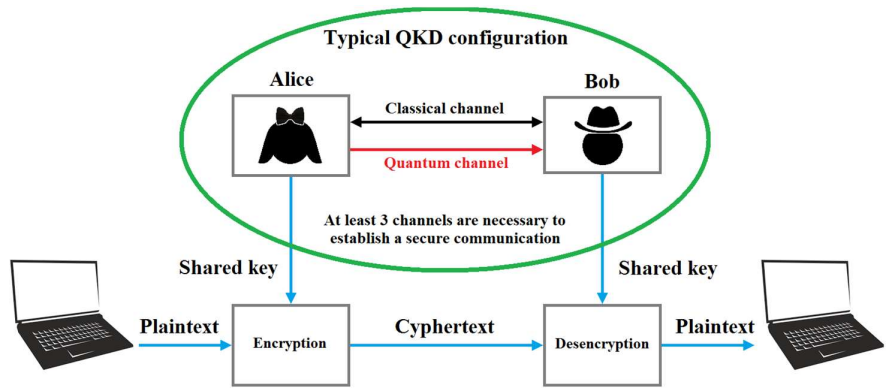

**Figure 1.** Typical QKD configuration with two channels for the distribution of the key, and an extra channel (public) for the transmission of the encrypted message.

In the rest of this section, we will refer frequently to Figure 1 as the natural environment where several of the implementation problems that this technology entails are highlighted.

Next, the different implementation problems of all QKD protocols are described according to the scope used for the distribution of the keys (land-to-land, space-to-land, land-to-space, or space-to-space), and as a consequence of this, it is evident that a new technique is required that avoids the mentioned problems to make efficient any practical implementation of this type of protocols.

*2.1. Issues common to all QKD implementation*

QKD protocols encode classical information (bits) in quantum states prepared on different discrete degrees of freedom of physical systems. The first and most important are those based on photons polarized in orthogonal states [24], entangled photons [25], and photons polarized in non-orthogonal states [26].

The security of these protocols is determined from the careful study of the effects of the intervention of an eavesdropper. The objective of any security study is to minimize the amount of information that an eavesdropper can obtain when attacking a certain scheme, knowing the value of some conspicuous metric that allows evaluating such information exposure, e.g., the Quantum Bit Error Rate (QBER), which is estimated during the distribution of the key and the conditions in which the experiment is carried out. From this study, the proportion of maximum secure key bits that can be extracted with classical post-processing of the distilled key is estimated. Although the security of the QKD seeks to be demonstrated in the most general scenario possible, in most analyses, different restrictions on the technological capabilities of the eavesdropper are assumed. Although this type of study only guarantees safety in a limited context, the results obtained help to understand and advance in a general demonstration of safety for different protocols. The result of these studies may yield new levels for the secure key rate, or modifications to the protocols to make them robust against different attacks. In most cases, the ultimate proof of security is an open problem.

The National Security Agency (NSA) [27] highlights the following technical limitations of the QKD protocols:
- Quantum key distribution is only a partial solution,
- Quantum key distribution requires special-purpose equipment,
- Quantum key distribution increases infrastructure costs and insider threat risks,
- Securing and validating quantum key distribution is a significant challenge, and
- Quantum key distribution increases the risk of denial of service.

Other also recent studies [28] tell us about a series of needs to take into account when working with these protocols such as privacy amplification, and information reconciliation [29].

However, the problems do not end there, in fact, they are just beginning. As we can see in Figure 1, the key distribution uses two channels [29], one classical and one quantum. These channels must be authenticated, since, although quantum communication attains the ability to detect the bugging and stop the communication, the key still needs to be passed to each other in a classical communication way to ensure that the subjects do not communicate with someone else. [30-32]. It is because of this apocryphal presence in the channels that configurations of several states [33] are used to create ambiguities that make it difficult for the hacker or eavesdropper to act.

The problems of attacks on the QKD protocols (e.g., man-in-the-middle attack, and some other particular cases of each type of implementation that we will see in due course) force a recurring distillation of the key to ensuring that it is not shared with a hacker. However, the distillation has a consequence that the key becomes smaller and smaller, in fact, a much smaller size than intended. This happens equally in terrestrial and satellite implementations of QKD. This forces the QKD protocol to be restarted again to obtain a final key, made up of the accumulation of the parts obtained in the different attempts so that at the end a key of the intended size is available.

Added to the aforementioned problems during a typical QKD implementation is the need to resort to:

- immense amounts of quantum memory, to hold a state at one point while its counterpart arrives at another point,
- time synchronization so that all the operations carried out by the protocol respect the correct sequence that it requires, and
- so on.

*2.2. Issues with terrestrial QKD implementation*

As mentioned above, the unconditional security of QKD systems is based on a thorough characterization of both the theoretical protocol and its experimental implementation. The latter is one of the most complex and constantly evolving tasks in the area. Recent protocols seek to minimize the effect of the imperfections of the experimental devices used. On the other hand, shortcomings or characteristics of the protocols and their practical implementations that can be exploited to violate their security are also sought [34-36].

Moreover, there are attacks particularly concentrated on imperfections of the single-photon detectors, where this type of action seeks either to manipulate the detection efficiency or to take advantage of some variation of this quantity in favor of the eavesdropper. On the one hand, there are attacks on detectors that work with detection windows. In general, the detection efficiency at the edges of windows does not go from 0 to the maximum value instantly. An eavesdropper can use this feature by selectively adding delays to signals intercepted in an attack [37]. Attacks that make use of detector dead-time to manipulate measurement results have also been proposed [38]. Other attacks are carried out by damaging detectors with very powerful pulses of light. Once damaged, the eavesdropper can control the detection signals with intense light (putting the detector in photocurrent mode) and thus perform an intercept-and-forward attack without increasing the QBER [39].

Some attack proposals focus on devices used to control quantum states, both for preparation and measurement. One type of attack, known as a "Trojan horse" [40], is based on introducing intense light (generally of a different wavelength than the one used for the protocol) into Alice's (or Bob's) station to later obtain information from the primed states (or the detection basis choices) from variations in the light reflected the quantum channel by the devices of the experimental setup or detection scheme (intensity, phase, etc. modulators). Another feature of state preparation devices that can be exploited by an eavesdropper is the wavelength dependence of their operation. For example, there are attacks based on beam splitters whose transmission factor (reflection) strongly depends on the energy of the incident photons [41]. An eavesdropper can intervene by sending signals with different wavelengths and thus force detection on a measurement or projection basis to a particular state of the protocol.

In general, for each of these attacks, there is an easy-to-implement preventative countermeasure, such as measuring the intensity of light entering Alice and Bob, interference filters, or temporal filtering of the detected signals. That is to say, the real importance of this area is to study what characteristics of the systems can be attacked and to propose a simple way to avoid said vulnerabilities, keeping the requirements on practical QKD systems to a minimum.

To the aforementioned attacks, we must add those specifically aimed at a particular protocol, such as the individual attack on BB84 (intercept and resend) [42], the optimal individual attack on BB84 [43] (cloning machine [44]), and the photon number splitting attacks [45-48]. Finally, Lucamarini *et al*. provide us with a complete list of attacks [49].

On the other hand, it's not just attacks that are the problems facing any terrestrial implementation of a QKD protocol, given that an implementation of this type presents a series of challenges that are difficult to manage in practice [50-58].

These challenges are presented in the form of very high technical requirements. Among the most outstanding, we can mention:

- Every implementation of QKD demands low channel noise [59]: Unfortunately, current implementations of QKD systems show relatively low-key rates, demand low channel noise, and use ad hoc devices. Every QKD protocol is subject to channel noise, which strongly conditions its performance and therefore the quality of key distribution.
- These configurations require authentication [60] of the channels used in key distribution, i.e., classical and quantum channels present in Figure 1.
- Its fiber optic installations have requirements for specific platforms and layers [61], i.e., it is not just any optical network, although this is strongly conditioned by the type of photons with which the protocol works (polarized or entangled).
- This requires synchronization [62] of both ends and recipients of the key.
- The network is too exposed to fake users, hence the previous mention about the need for channel authentication [63].
- All protocols need a key distillation process [64] to achieve a certain degree of security about the privacy of the distributed key, although as we have mentioned this process reduces the size of the key so much that the protocol itself must be applied recursively until the preset key size is reached.
- Fiber optic cabling for terrestrial implementations of QKD requires quantum repeaters every certain number of kilometers [65], which in turn requires a large amount of quantum memory. The problem is that the key is exposed in its passage through them. There are currently two well-defined lines of research, the first has to do with the development of quantum repeaters that do not require quantum memory, at least not that much, and the second is to replace the same quantum repeaters with some type of implementation based on quantum teleportation [66].

*2.3. Issues with space-to-ground and ground-to-space QKD implementation*

To the common problems of any implementation of a QKD protocol, we must add those that arise when including a satellite that transmits photons to Earth into the equation. As a direct consequence of this, the so-called beam effects appear [67], which generate low-performance problems [68]. Consequently, the protocols are subject to transmittance problems [69], that is, the photons that are emitted by the satellite toward the Earth cannot compete with the sun when its beams reach the lens of the optical ground station (OGS) telescope, therefore, the system can only be operational at night, as long as it is a clear night (without clouds) and preferably the OGS is located in an area with a low level of atmospheric pollution.

The atmospheric effects make it necessary to take into account the ephemeris and climatic considerations with strong seasonal and time limitations, which is essential for the modestly successful delivery of the key [67]. In addition, geographical aspects must be considered, for example, in Saudi Arabia, there are five types of dust, which means that in the event of a storm, the OGS must have positive pressure so as not to let the dust enter the delicate optical instruments, even when the OGS radome is closed.

Some of these problems were faced by the Chinese mission Quantum Experiments at Space Scale (QUESS) or Micius [70] between 2016 and 2018. In cases like the previous one, to all the difficulties mentioned we must add the use of a double optic with an object of distributing entangled photons to two points on Earth at the same time, which forces several demanding design decisions:

- a higher orbit vs. an increase of the angle between both optics, due to the geometry of the simultaneous ground focusing, and
- a double optics vs. simple optics, although in the latter case at the cost of a lot of quantum memory.

This strongly conditions the type of experiments that can be carried out with the platform, because if the satellite has double optics then it is possible to carry out teleportations between two points on Earth while the satellite is a mere distributor of entangled photons. On the other hand, if the satellite has a single optic, it is more convenient to carry out teleportations between a point on Earth and the satellite so that the quantum memory is found in the OGS. In that case, it is the OGS that emits an entangled photon of each pair to the satellite and keeps the others. Teleportation in the opposite direction would imply that the satellite would emit one photon of each entangled pair to Earth and keep the other in onboard quantum memory, with what this implies in cost, size, and complexity of the platform.

On the other hand, like their terrestrial counterparts, QKD protocol implementations thanks to the use of satellites deal with the need to use classical and quantum channels [71]. Moreover, we must consider the impossibility of distributing the keys underwater [72] in the case of submerged submarines, the problems associated with different configurations of satellite repeaters [73], and severe climatic factors such as climate change and associated droughts, which involve smoke caused by fires that require the use of intermediate relays between the satellite and the OGS through the intervention of drones [74, 75]. These operational difficulties of the satellites involved in QKD protocols have given rise to a recent trend consisting of replacing the satellites directly with drones, which, by flying much closer to the OGS, avoid several of the problems mentioned above [76].

As mentioned above, all QKD satellites share many problems with strictly terrestrial setups. One of the most complicated lies in the small final size of the key after successive distillations to ensure the privacy of the key, i.e., only reserved for Alice and Bob, and no one else. To reach the pre-established key size, successive applications of the protocol are required, which in this context translates into successive orbits of the satellite, taking into account all the drawbacks mentioned above. However, a QKD satellite has serious problems of its own [54, 77, 78]. Among these problems, the need for synchronization of the protocol between both recipient points of the key distribution through the Global Positioning System (GPS) stands out. This means that if the QKD configuration is used during a conflict by the government or the armed forces and the GPS satellite network is intervened, interrupted, or attacked by the adversary, the QKD satellite procedure is automatically out of service.

We must also take into account a serious problem that affects the link between the satellite and the OGS, or every QKD satellite itself, and that is satellite espionage, i.e., problems affecting the physical security of the satellite. In this regard we must mention:

- man-in-the-middle attacks that try to intervene in the link between the satellite and the OGS [79],
- the existence of tarantula type satellites with prehensile extremities [80], which would allow the QKD satellite to be physically seized, thereby automatically interrupting the service, and which would lead to the acquisition by the adversary of the key history and part of the QKD protocol with its eventual innovations in that particular implementation,
- the presence of passive spy satellites [81], which are positioned in the proximity of the QKD satellite to listen to any signal coming from it, and
- the presence of active spy satellites [82], which try to stealthily modify (emulating a failure) the satellite's telemetry and orbit so that it is lost or incinerated by re-entering the atmosphere.

Finally, we must mention the damage caused by the presence of so many satellites around the Earth, some in configurations of 300 or more units, generally in Low Earth Orbit (LEO) at approximately 500 km altitude, i.e., in this context the satellites are the aggressors.

There are more than 8,000 satellites in orbit and countless pieces of space debris surround the planet [83-88]. The sky has become a veritable space dump. Therefore, it is not reasonable to continue to orbit constellations of hundreds of smallsats (CubeSats, PicoSats, and so on) in LEO, as this impedes the work of astronomers [89-93], who track asteroids, meteors, and comets on a collision course with Earth. In other words, the proliferation of the aforementioned satellite constellations conspires with the achievement of life on Earth as we know it. In addition, with this trend, one day soon Earth-based astronomy will die, so this important scientific discipline can only be carried out from orbiting platforms such as the Hubble Space Telescope, Web, and those that will inevitably have to be built soon. International Astronomical Union launches new center to fight satellite mega-constellations threat [94]. This new center will coordinate work to mitigate the effect of satellite constellations on astronomy [95]. The main idea is the protection of the dark and quiet sky from satellite constellation interference [96]. SKA observatory and international partners petition Union Nations for the protection of Earth's dark and quiet skies [97].

Finally, given that satellites generally have a useful life, on average, of two to three years, a constant renewal of the satellite fleet of a mega-constellation is regularly required or else resort to a no less expensive method, consisting in orbiting robots, which could help fix and fuel satellites in space, i.e., machines will soon have a go at maintaining a fleet of small space-craft orbiting Earth [98]. In other words, more satellites to assist other satellites, which would further aggravate the problem.

Although we have the vague feeling that this uncontrolled trend will one day cover the sun, one thing is certain, at this rate, it will not be long before the aforementioned constellations affect the departure of ships into space.

*2.4. It is then required*

Based on what has been stated in this section, we conclude that to achieve a correct QBER level and overcome the difficulties of both QKD configurations (terrestrial and satellite), a system that shares keys must have the following attributes:

. It should not use satellites for QKD due to its problems, i.e., we establish a non-satellite criterion, where we avoid the use of additional satellites for QKD, even though fog cloud is also a satellite,

- This does not require synchronization by GPS satellite constellation,

- This does not require authentication of the channels, in fact,

- This does not require quantum or classical channels,

- This does not require the distillation of keys,

- It does not require quantum memory,

- This does not require quantum repeaters,

- This does not have limitations of time, season, weather, or geographical conditions to share the key, and

- It is not subject to the action of hackers or eavesdroppers.

Therefore, if the current implementation of QKD configurations has so many complications, it is better to resort to a secure configuration, that is, one that incorporates the attributes mentioned above, and since it is about security, we consider that the technology proposed below (Non-distributable key sharing: NDKS) is the best option when it comes to sharing keys securely.

## 3. Non-distributable key sharing (NDKS)

In addition to incorporating all the virtues exposed in Subsection 2.4, this technique must comply with a series of requirements when it comes to replacing the expensive, inefficient, and complicated terrestrial and space configurations of QKD. These requirements are:

- This technique must be able to be implemented in boxes, called NDKS units, which, when separated, can share a common symmetric key for encoding and decoding messages, in the sender and receiver, respectively. So that at the time of installation both boxes start with a few common parameters and operators, which can be renewed (and which will be defined later in this section) as many times as desired, through secure transmission, where security is provided for the same boxes. NDKS units have the particularity of being able to share the same key regardless of the distance that separates them and without using any classical or quantum channel for this purpose. They can also work in a coordinated or synchronized way in configurations of more than two boxes, in fact, for any number of them.

- NDKS units act as if they are under the influence of a kind of virtual entanglement. If it were more than two NDKS units, then we would speak of a multi-virtual entanglement since it is as if the boxes shared an underlying reality, and when asking the boxes for a key (measurement), both would respond with a common result. (i.e., the collapse of a common virtual wave function). This is something similar to what happens when the Bell state $|B_{00}\rangle = (|00\rangle + |11\rangle)/\sqrt{2}$ is measured for two NDKS units [99], or a Greenberger–Horne–Zeilinger state, for the case of 3 or more NDKS units virtually entangled, that is, $|GHZ_n\rangle$ [99]. In other words, the NDKS units reproduce or emulate a putative behavior of quantum mechanics [9] even though they are strictly classical entities. After all, even in QKD, the shared key to be used is ultimately digital (classical).

- A configuration based on NDKS units should create a *caveat emptor* (information asymmetry). That is, a key easily shareable by the two or more NDKS units, but impossible to acquire or deduce by the eavesdropper. This implies the existence of a significant Entropy gradient between the NDKS units and the eavesdropper, with Entropy equal to 0 between the NDKS units, and Entropy equal to 1 for the eavesdropper. After all, the important thing is not that the key is more or less random, but that the adversary does not know it and cannot deduce it. Consequently, a configuration based on NDKS units does not require a chip for quantum random number generation (QRNG) [100] for its operation.

- An NDKS unit-based configuration uses sequential synchronization instead of time synchronization. In this way, even in the event of a failure of the channel that transmits the encrypted message, all the NDKS units maintain the order while waiting for the restoration of the service, keeping an exhaustive count of the sequence, and thus the lost message can be collated with the same key. For this, buffers and a frame composed of header and data are used, the same as the frames used by IP datagrams. In this way, the communication can be asynchronous without altering the reconstruction of the total message. This positions the NDKS technology as an ideal tool to be used in Blockchain operations [101]. Moreover, the protocol itself is the architect of synchronization and order in bidirectional communication. In this way, the operation of a configuration based on NDKS units does not require the clock of a constellation such as GPS for its synchronization, which is extremely convenient in the event of a possible attack or hacking of said satellite network in the event of a conflict.

- Taking into account that in its simplest operation each NDKS unit is directly associated with a computer (generally through a USB port), they must be able to act in two well-defined modes:

a) the sending computer interrogates its NDKS unit and it answers with a key. The computer can use the key to encrypt the message using Vernam's [102] ciphered based on XOR, or another. On the other side of the channel, the receiving computer interrogates its NDKS unit, which replies with the same key as the sender, then the computer will use that key to decrypt the message, and

b) the sending computer sends the message to be encrypted to its NDKS unit, which returns it encrypted. The computer then transmits it over the channel to the receiving computer, which sends it to its own NDKS unit for decryption.

- The NDKS unit system does not preserve the link between the sender and receiver, nor the sender's or receiver's computers, not even the original message information, but only the encrypted message information, that is, the NDKS unit system is not a firewall, neither an antivirus software, nor an intrusion monitor, but simply a technology to encapsulate Information. In this regard, the NDKS drive system is 100% responsible for the integrity and security of the encapsulated (encrypted) data. But only about it.

- The NDKS units are essentially composed of nonlinear operators (they do not comply with the superposition principle), parameters, and independent and dependent variables. These operators will reproduce the same binary sequences if the parameters and variables have the same values. That is, it is a deterministic process between the NDKS units, but of apparent absolute randomness for the eavesdropper, which only sees a non-stationary and indecipherable sequence of zeros and ones, which constitute the encrypted message. This is what we referred to in a previous requirement when we mentioned the important entropic gradient.

- The nonlinear operators mentioned above have a very high sensitivity to the parameters and variables that compose them, i.e., a small change in one decimal of the parameters or variables will give rise to a completely different binary sequence (shared key).

- A cryptographic system based on NDKS units gives rise to dynamic encryption whereby each message and each symbol of the same message can be encrypted using a different key each time. In this way, we can encrypt the message using Vernam's [102] ciphered based on XOR, or another. Even if the message always consists of the same symbol, the NDKS units will assign it a completely different encoding each time, breaking the periodicity of the message. In this way, this cryptographic system is immune from any brute force attack, statistical analysis, pattern recognition, or disambiguation processes aimed at deciphering the key and with it the message. The statistic changes with each transmission, so the Entropy is practically fixed at one during the entire cryptographic process.

*3.1. Two contexts in which to apply NDKS units*

The first context in which the NDKS units can be applied has to do with a pure CyberSecurity process, that is, the protection of information and data integrity in a communication such as that represented in Figure 1. The plain text enters the Gateway, which is linked to the NDKS unit under either of the two working modes explained above, that is, the Gateway requests a key from the NDKS unit and with it encrypts the message, or sends the plain text to the NDKS unit and the cipher-text is returned to it.

As we have repeatedly mentioned before, the applied coding can be of the Vernam [102] type or any other that the configuration designer considers more convenient.

Either of the two modes of use of the NDKS units used does not protect the plain text at the Gateway before the encryption process (at the sender) and after the decryption process at the receiver's Gateway. For this reason, the configuration must necessarily have firewalls on both sides of the channel to protect plain-tex while it is present.

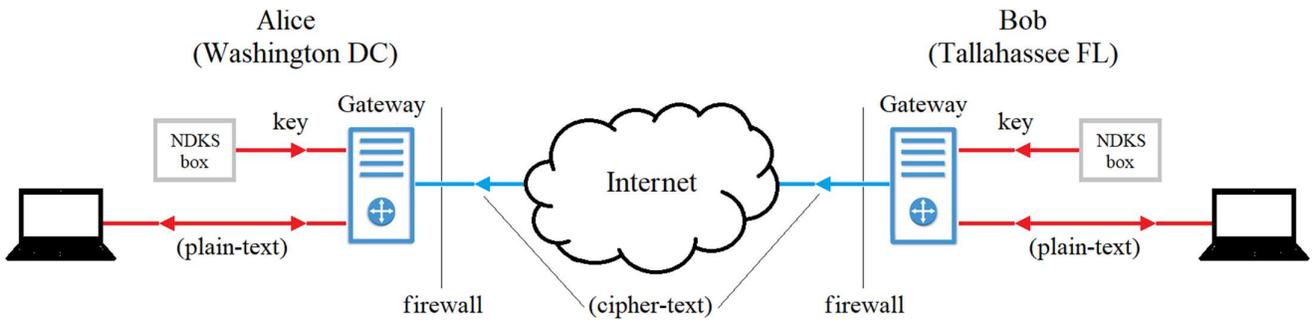

**Figure 2.** NDKS units in a CyberSecurity context.

The second context alludes to the use of NDKS units in the particular case of an IoT configuration, such as the one in Figure 3, where the Gateways are specifically linked through a fog cloud. Figure 3(a) encrypts the plain signals coming from all the devices connected to this configuration and of which real-time monitoring is desired, as well as their control, while Figure 3(b) shows they encrypt the plain commands coming from the user's computer to condition some parameter or variable of the aforementioned devices: Revolutions per minute, temperature, humidity, light, and so on.

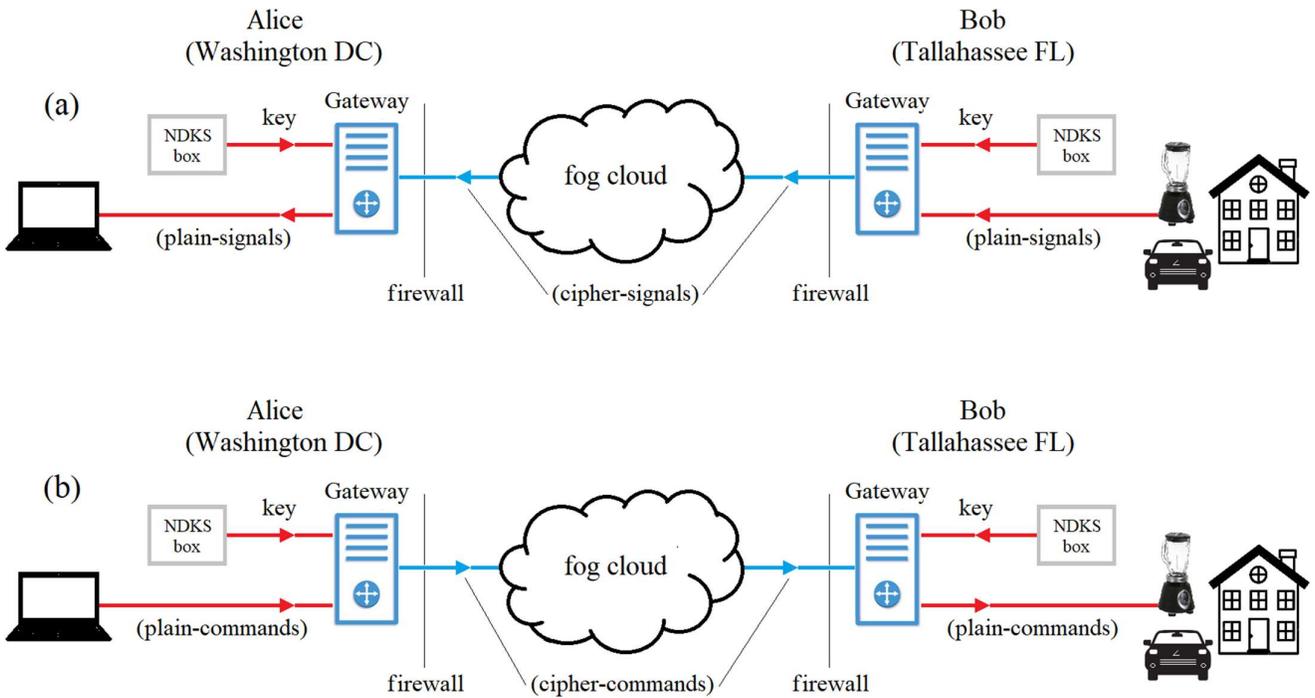

**Figure 3.** NDKS units in an IoT context: a) Sensing, and b) Control.

Both monitoring and control can be carried out under the real-time, batch, and semi-batch modes, to use the fog cloud when the connection service or the contracted rate is cheaper. In the case that concerns the physical security of tangibles, the real-time modality prevails over the other two, the same as in the case that sensed and controlled devices have to do with the field of defense and security since wrong or delayed command can trigger a conflict.

*3.2. NDKS protocol*

As we mentioned before, the NDKS units are based on a set of non-linear operators, which do not comply with the superposition principle, that is:

$$T(x+y) \neq T(x) + T(y), \text{ and} \tag{1a}$$

$$T(\alpha x) \neq \alpha T(x), \tag{1b}$$

where $T$ is the non-linear operator, $x$ and $y$ are the independent variables, while $\alpha$ is a parameter (constant). These operators are used in a composition of functions:

$$(g \circ f)(x) = g(f(x)), \tag{2}$$

where $D(g \circ f) = \{x \in D_f \,/\, f(x) \in D_g\}$, and $D(\bullet)$ means *domain of* "$\bullet$".

Then, the internal model of the NDKS units consists of a non-linear multidimensional system, which in its most generic form is:

$$y_1(k_1) = f_1(y_0, p_{1,1}, p_{1,2}, \ldots, p_{1,n_1}, k_1), \tag{3a}$$

$$y_2(k_2) = f_2(y_1, p_{2,1}, p_{2,2}, \ldots, p_{2,n_2}, k_2), \tag{3b}$$

$$\cdots$$

$$y_N(k_N) = f_N(y_{N-1}, p_{N,1}, p_{N,2}, \ldots, p_{N,n_N}, k_N). \tag{3N}$$

The composite function of this system of Equation (3) is:

$$(f_1 \circ f_2 \circ \cdots \circ f_N)(k_1, k_2, \ldots, k_N, p_{1,1}, \ldots, p_{N,n_N}), \tag{4}$$

where $y_i$ are the dependent variables $\forall i$, with $y_0 = 0$, $p_{i,j}$ are the parameters of the system $\forall i, j$, and the $k_i$ are the indices (independent variables) $\forall i$. A more practical case to be used in the NDKS units with the highest possible efficiency is:

$$y_1(k) = f_1(p_1, p_2, k), \tag{5a}$$

$$y_2(e) = f_2(y_1, e), \tag{5b}$$

$$y_3(e) = f_3(y_2, d), \tag{5c}$$

$$y_4(e) = f_4(y_3), \tag{5d}$$

where $p_1$ and $p_2$ are parameters, $k$ is the key index (independent variable), $y_1$, $y_2$, $y_3$, and $y_4$ are dependent variables, and $e$ is the index of the key element (independent variable). As we will see later, if $f_2$ is a function of the sinusoid type, then $f_1$ is its angular frequency dependent on the parameters $p_1$ and $p_2$, and the key index $k$. The composite function of this particular system is:

$$(f_1 \circ f_2 \circ f_3 \circ f_4)(k, e, d, p_1, p_2). \tag{6}$$

However, what does the independent variable $d$ represent? We will explain it with a concrete example. We generally use sinusoids in the case of the functions $f_1$ and $f_2$ because they are strongly non-linear functions and with a module less than or equal to 1, therefore:

$$y_1(k) = |p_2 \sin(p_1 k)|, \quad (7a)$$

$$y_2(e) = |\cos(y_1(k)e)|, \quad (7b)$$

$$y_3(e) = numtovec(round(10^d \, y_2(e))), \quad (7c)$$

$$y_4(e) = (1 + sign(y_3(e)/2 - floor(y_3(e)/2) - 1/4))/2, \quad (7d)$$

where:

$y_1$ is a vector / $k \in \mathbb{N}$, specifically, *for k = 1 to number_of_keys*,

$y_2$ is a vector / $e \in \mathbb{N}^0$, specifically, *for e = 0 to key_size-1*,

|•| is the absolute value of "•",

*sin*(•) means *sine of* "•",

*cos*(•) means *cosine of* "•",

$y_3$ is a vector of *key_size* digits / each digit $\in [0, 9]$,

*round*(•) means *round to the nearest integer of* "•",

*numtovec*(•) means *number* "•" *to vector*,

$y_4$ is a vector of *key_size* digits / each digit $\in [0, 1]$. $y_4$ represents each key,

*floor*(•) gives as output the greatest integer less than or equal to "•", and

*sign*(•) extracts the sign of a real number "•".

Through a numerical example, that is, by giving values to the system of Equation (7) we can better understand what *d* represents.

Suppose that $N = 10$, $e = 3$ (we must remember that *e* can be any integer value from the interval [0, $2^{N-1}$]), $p_1 = 0.52$, $p_2 = 25$, $k = 1$, and $d = 3$, thus:

$$y_1(1) = |25 \sin(0.52 * 1)| = 12.422003446093$$

That is, $y_1(1)$ is the angular frequency of $y_2(3)$.

$$y_2(3) = |\cos(12.422003446093 * 3)| = 0.907668453135977$$

$$y_3(3) = \begin{cases} 10^3 * 0.907668453135977 = 907.668453135977 \\ round(907.668453135977) = 908 \\ numtovec(908) = [9 \; 0 \; 8] \end{cases}$$

That is to say, *d* allows us to select the decimal with which to work.

$$y_4(3) = \begin{cases} lsd(908) = leo([9 \; 0 \; 8]) = 8 \\ (1 + sign(8/2 - floor(8/2) - 1/4))/2 = 0 \end{cases}$$

Where *lsd* means *less significative digit*, and *leo* means *last element of*, while the second line of $y_4(3)$ is equivalent to "*if 8 is even, then* $y_4(3) = 0$, *else* $y_4(3) = 1$".

Next, we generate some examples of keys to evaluate the high sensitivity of the system of Equation (7) with a slight variation of the parameters, which shows a great multiplier effect generating completely different keys with a small shift in the parameters.

For $N = 10$, *key_size* = $2^{10}$ = 1024, $k = 1$, $p_1 = 0.52$, $p_2 = 25$, and $d = 3$, the key is:
0010010100111110010000100111100110100010100101110100010010001100111101100111000000100111010011100111111111010100000101111001000011011000110110001101100111111111100000000101001011110010001000111100000001101110010101010001100010011001001000101101001001000110111100000100010001001100100101100001000011100100111001110011001111101000111010001110101011001000111110011010011100001110000000010

00110110111100000010001000001100111001111001100101011110011111001110010111010000111010110010101010010010000011011101100001001000101111011001011111000100011000101110011101100010010010010110100100111100011011010010110001011000011111000000001000100010000111100001111000111110111101010010001001101011000100111110001000010000111000100111010100001000101110011001000111001100100011010111001100011101000100011010100110110000111001111000101000001001101100100001100001110100000001100011100011111110101111011110101100000110111011000000110110100000000101110111100100000010000100101010000111100000001000101011111000111010010101100001001001110 1.

For $N = 10$, $key\_size = 2^{10} = 1024$, $k = 1$, $p_1 = 0.52001$, $p_2 = 25$, and $d = 3$, the key is:
00100111001101101100110101010101110110111100010101011000111001000001010001110000110011001000001000100000100001100011001111101000010011001011101100011110011001000111111001110001001101000011111100000010010000100111011000011011000110100111011111110011101111100010011110001111100010111011111011100100110100111011101001001011010010111111001101111101110100100001111010011101110001110110110011101000010011000110110010111011011010001010001001010111001000011111001010000011101000001001000001000110010101011011010011101001100111001111111101100111000001001110000101000111100000111010010110110111100100010110010100101101000110011101000101111100110111101010111000010110001010010001101100011000101010000110010001101010000101000011110011000101001010100110010010110100111111010001001000001010100101111000110000101100110101110011000011011110111011111110000110100001100000101000100110101000100111011100011101110000000001100001000101111000111100001001010110110000010000111000110000011011111101011011100001010101110001110000001101

For $N = 10$, $key\_size = 2^{10} = 1024$, $k = 2$, $p_1 = 0.52$, $p_2 = 25$, and $d = 3$, the key is:
01101000101100110010010100011000111101111011100101100101111011111110110001000011111110001001110001011001111010010100101010011001001101010011111100101011000110110011111111110010011010001111111010010000101111111100101111001111011111111011101001101100000011010100110010010001010011011100101010010111010100001100110111111101010011001010100101100101111100100000001101010011000000001000101011101001010010000001010010100001111000110011001111110100010001110011011101110100001010001111011011100111011110001001001001101100111000110111111111101010111000100101001010101111110100110100000000001000000100100010001000000000001011110010000011010101000001010001101100011100011001011110001101100111111100111101100010111110010011010001110000101001011111111101010001000110100111001100010110010110001100011110000000100100001100111001101000000100011010000011101000111010010001101000111000111100101000101100110010110010001001100100000001000011101001010000111010001110100000000110010010110011110000100110001011001001100010111010000110011110001111 1.

These were just a few simple examples that show the sensitivity to the parameters that the operators used have, in such a way that changing the decimals of said parameters a little completely changes the binary sequence, that is, the key.

Other nonlinear functions that can be applied as operators instead of the sinusoidal ones of the previous numerical examples can be polynomial, hyperbolic, and so on type functions.

Specifically, the nonlinearity of the operators is due to two reasons:
- the aforementioned amplification factor, and
- simple functions generate non-ergodic binary sequences.

Finally, with these examples, we have verified most of the slogans raised above.

**4. Discussion**

In this section, we will develop two important topics related to this technology:

- the testing and homologation protocol, without whose approval this technology cannot be commercialized, and
- the context analysis, whose purpose is to compare this technology with others of quantum cryptography, in general, and QKD, in particular, and finally give it ubiquity from as many points of view as possible, e.g., economic, financial, manpower necessary to its implementation, and so on.

*4.1. Testing and homologation protocol*

The testing and homologation protocol of the NDKS system units consists of two main stages:

- No-Channel key sharing, and
- Encrypted communication via regular channels.

4.1.1. No-Channel key sharing:

With both laptops in Figure 4 completely disconnected from wifi, Bluetooth, and any type of cable that connects them to a LAN, WAN, or Internet nets, both laptops interrogate their respective NDKS unit about a key. That is, the key is exchanged instantly without any channel intervention. The Faraday's cages prevent the intervention of any external channel, such as wifi, Bluetooth, and so on. In other words, both laptops can only share the same key exclusively using the NDKS units. Then, the sending laptop encrypts the original message with the key obtained from its NDKS drive and deletes the original message, keeping only the encrypted message.

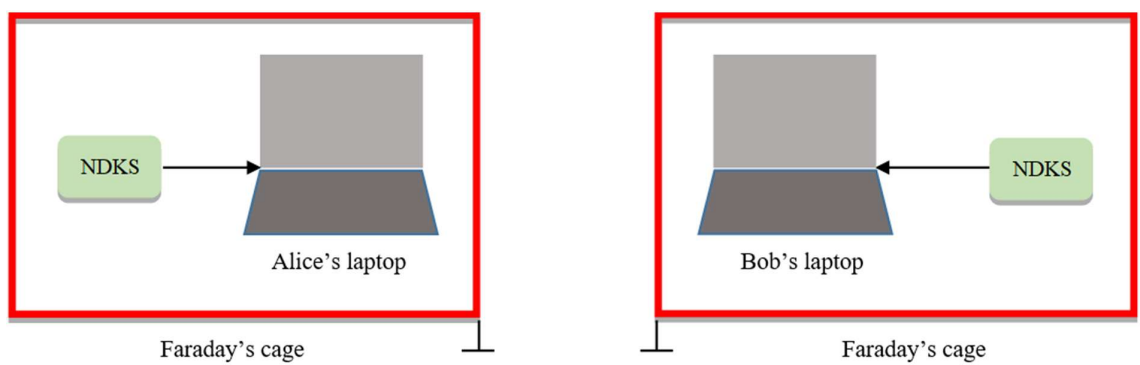

**Figure 4.** No-Channel key sharing.

4.1.2. Encrypted communication via regular channels:

Both laptops get rid of their respective Faraday cages and then connect to the same channel (e.g., the Wi-Fi network of the office where the test is performed). In this instance, hackers also connect to the same network, which will try to decipher the encrypted message that will be transmitted. Thus the hackathon begins. The sending laptop transmits the encrypted message. Once the receiving laptop receives the complete encrypted message, it disconnects from the Wi-Fi network, Bluetooth, and all types of network cables that connect it to a LAN, WAN, Internet, etc. The receiving laptop then decodes the encrypted message using the key provided by its NDKS unit.

A recognized and independent expert verifies that the message originally written by the sender (Alice) from her laptop matches the one decoded by the recipient's (Bob) laptop. The expert questions the hackers about the message that they may have decoded during the transmission of the encrypted message. The expert renders his verdict.

The roles of Alice as sender and Bob as receiver are permanently swapped seamlessly and without the intervention of any special or additional switching system. The protocol is completely bidirectional. An NDKS configuration is a symmetric key technology that facilitates all operations related to bidirectionality.

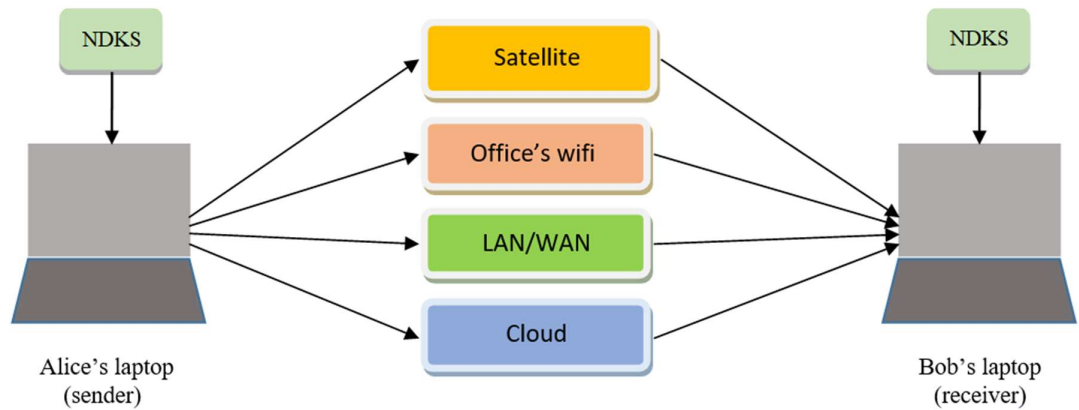

**Figure 5.** Encrypted communication via regular channels.

If the result of the hackathon is successful (positive verdict), the NDKS units are approved for subsequent sale. Finally, this protocol and, in consequence, the NDKS units have been designed taking into account the following considerations:

- they can be connected to the laptop through one of its external USB ports, or internally in any slot of it, although said slot should not have any type of port that links it to the outside,
- the information that goes out and enters the NDKS units is organized in a frame structure with header and data, which is encrypted,
- if one of the NDKS units is stolen, it will not be able to connect to a new laptop since the NDKS units share a sync word within the frame structure, which results from a recognition procedure of the physical characteristics of the laptop from which was stolen, as well as the characteristics of the network where, said the laptop was,
- the aforementioned synchronism word is renewed in each transmission and is encrypted within the frame,
- if an NDKS unit is opened, its contents will automatically be erased, so these units are built like a bomb,
- in any network of NDKS units, one of them will act as master and the rest as slaves, where these roles can be changed automatically at any time in a transparent way to the users, if the master is stolen, opened, or destroyed,
- any new NDKS unit can be incorporated into the network at any time by simply requesting permission from the NDKS unit network administrator, privately sending through an alternative channel the physical data of the laptop and the characteristics of the network in which the latter will be connected, then the administrator will consult with the competent authority if said incorporation is scheduled. In a positive case, the master will send a synchronization signal to the entire network within the encrypted frame to put them all in an entanglement state,

- this technology should prevent the government from applying an "Internet kill switch" in case the network of networks is attacked,
- NDKS units use dynamic encryption, where the key can change even for each symbol of the same message,
- regularly it can be verified by an alternative channel if Bob decrypts what Alice encrypted, and employs a confinement procedure whereby all the units of the network are gathered in the same place, to discern if an eventual error found is due to the NDKS units or due to noise on the transmission channel,
- the keys generated by the NDKS units must not show regular patterns. This was verified by the developed examples,
- the NDKS unit that transmits is the one that generates and emits a synchronization signal inside the encrypted frame,
- the configuration can also be programmed so that the network master of the NDKS units is the unit that transmits, for which the administrator role is in a permanent relay state unless the network is used to collect data from a device with no intention of emitting any control signals, that is, a pure and exclusive monitoring state,
- NDKS units recognize three possible states at a low level: 0 (zero), 1 (one), and nothing (absence of information), so that when nothing is transmitted, pure coded zeros are not used, because in that case if the encryption is of the Vernam type, keys that are not used to encrypt useful information would travel through the channel, that is, keys would be wasted unnecessarily,
- NDKS units can use any type of encryption technique, in such a way that if we want the key to having the same length in bits as the message to be encrypted, then we can use Vernam's ciphered (XOR), instead, if the key is of smaller length in bits than the message to be encrypted, a unifying file will then be used, which can be the result of a folder containing all kinds of audio, image, video, and text files, and
- as the operators used by the NDKS units are non-linear, the eavesdropper sees the keys as having random characteristics and resulting from an apparent probabilistic or stochastic source. This happens because the hacker does not understand (i.e., does not know) the underlying law behind the generation of the key, but we know that it is a 100% reproducible process (because otherwise, the network of NDKS units would not work), due to we know the generating law. So, who are the keys random for? It is evident that if we do not know the laws that govern the universe, its behavior will be indecipherable to us, on the other hand, if we know them, it will be harmonic and elegant. This is what we mean when we talk about caveat emptor (information asymmetry).

*4.2. Context analysis*

In this subsection, two comparative tables are presented to give ubiquity in the current international context to the technology of the NDKS units. Table 1 presents a comparison between NDKS technology versus ground and space versions of QKD. The evidence contained in this table is overwhelming in favor of the technology of the NDKS units from every point of view, i.e., technological, financial, etc. Moreover, Table 2 consists of a comparison between NDKS technology and the Chinese mission QUESS/Micius [103]. In the penultimate row of Table 1, we mention an especially important item, the "legal procedures", among which we mention certain tests. Those tests are required by the launcher as a *sine-qua-non* condition to embark on the satellite unit. These tests are:

- *Vibration testing on a shaker*: This test tries to reproduce the vibrations that the satellite will suffer at the time of liftoff from the launcher.
- *Calibration and testing*: This test refers to the evaluation of electronics and optical components, batteries, etc.

**Table 1.** Comparison between NDKS technology and the ground and space versions of QKD.

| Attribute | Ground QKD | Space QKD | NDKS |
|---|---|---|---|
| Dynamic vs static coding[1] | static | static | dynamic |
| Key size in bits | small after distillation (less than desired) | small after distillation (less than desired) | unlimited |
| Control over final key size | No | No | Yes |
| Number of keys generated during each operating cycle | less than one[2] | less than one | unlimited |
| Infrastructure need | Yes | Yes | No |
| Manpower | Highly qualified | Highly qualified | Low skilled |
| Maintenance need | Yes | Yes | No |
| Startup cost | Very high | Very high | Null |
| Legal procedures[3] | Too many | Too many | None |
| Contractors need | Yes | Yes | No |

1. This point refers to working with a single key or if it changes with each message or symbol of each message as the NDKS units do.
2. This is because the distillation process (to ensure that the eavesdropper does not share portions of the final key with us) of the key leaves the key smaller than desired.
3. This implies building permits (channeling and fiber optic laying), tests (in the case of the space version) without which the satellite cannot be uploaded to the launcher, etc.

**Table 2.** Comparison between NDKS technology and the Chinese mission QUESS/Micius.

| Parameters | QUESS/Micius [103] | NDKS units |
|---|---|---|
| Origin | China | USA |
| Construction time | 10 years | less than an hour |
| Weight | 631 kg | 250 g |
| Useful life | 2 years | *in eternum* |
| Orbit | 488-584 km | 0 km |
| Launch year | 2016 | 2022 |
| Primal key[4] | 300,000 bits | 192,000,000 bits |
| Efficiency[5] | At night, without clouds and with low pollution (much less than 100%) | 100% |
| Final cost | $ 100 M | $ 34/unit |

4. NDKS units can generate even larger keys and deliver them complete or in single bits (as streaming of ones and zeros) in order not to saturate the internal memory of these units.
5. This refers to the impossibility of competing with the sun and the transmittance due to atmospheric and meteorological conditions.

- *Vacuum and thermal cycling test*: This test studies the response of the satellite and its components against a thermal gradient in space with more than 200 ºC of thermal amplitude, which happens when the sun is absent and present at those times as a consequence of the orbit of the satellite around the Earth.
- *CubeSat attitude determination and Helmholtz cage design*: A Helmholtz cage is designed, built, and tested to provide a dynamic, 3-axis, uniform magnetic field to cancel the Earth's magnetic field and create an environment similar to the geomagnetic field a satellite would experience on-orbit.
- *Impact test*: Esta prueba atañe a la respuesta mecánica del satélite y sus componentes.
- *Radio-frequency (RF), electromagnetic compatibility (EMI/EMC), or anechoic chamber*: EMC tests such as Radiated Emission and Radiated Immunity are applied in a room with isolated electromagnetic signals. EMC (or anechoic) chambers are designed to create an enclosure with an extremely high level of shielding attenuation against electromagnetic interference. The idea is to study the radiation emitted by and/or radiated immunity of the satellite.

Comparisons complementary to those of the tables of an extremely heterogeneous nature are:

- *Launch cost*: Millions of dollars (due to the size and weight of the Micius satellite) vs. $0 for NDKS units,
- *Number of satellites*: Micius one to date, although it has not been in service since 2018 (2 more are planned) vs. none by NDKS technology, and
- *Required Manpower*: Highly Skilled for Micius, while in the case of NDKS units, these can be assembled by K10 students.

Moreover, although similar products are promoted for key delivery [104], which, however, are based on QKD technology, we must understand that, unlike the NDKS units, the aforementioned products require the typical channels of a QKD protocol to carry out the key sharing, that is, they need a classical channel and a quantum channel to be able to carry out the distribution of the aforementioned keys.

Another outstanding aspect of NDKS technology is that it is fully compatible with internationally established CyberSecurity standards, even in the most specific aspects, such as the ISO/IEC 27001 Standard: Information Security Management [105], or that of the International Association of Classification Societies (IACS), specifically regarding the recommendation on cyber resilience [106].

On the other hand, no matter how spectacular laser links are tested between satellites, airplanes, ships, and ground stations [107], these links can only be effective at night, with good weather and low pollution levels, unless the link is used only between satellites or between the same and a spacecraft leaving the Earth's atmosphere. These problems are largely overcome by the technology of the NDKS units.

Finally, as it is a dynamic encryption technology, that is, where the key changes with each message or with each symbol of the same, the NDKS units are completely immune from attacks such as those that can break an 8-character password in less than an hour [108]. Even in the case of 8-bit passwords, since it is dynamic encryption, the NDKS units make the keys change with each symbol of the message, making it impossible for a hacker to act.

**5. Conclusions**

The technology provided by the NDKS units has shown to dispense with the main elements that make the implementations of the QKD protocols work, that is, the technology presented in this work:

- This does not require a QKD satellite with all its problems,
- This does not require synchronization via a GPS satellite network,
- This does not require authentication of the channels for the delivery of the keys, since it simply does not use them for that purpose,
- This requires no key distilling, so the key is always the expected size even on the first try,
- It does not require quantum memory,
- This does not require quantum repeaters, so the key is never exposed during its passage through them as it happens in terrestrial implementations of QKD,
- This does not require a quantum or classical channel for key distribution,
- It does not have time limitations to share the key, nor do they depend on the weather or the pollution of the environment,
- It is not subject to the action of hackers or eavesdroppers, and,
- This does not depend on many other factors without which a cryptographic implementation based on QKD could never work.

Reasons like these, and many others, make the technology of the NDKS units the chosen one when it comes to providing maximum security to sensitive communication channels (Government, Defense, Fintech, and so on) compared to their competitors, that is to say, terrestrial and satellite QKD implementations.

**Author Contributions:** M.M. conceived the idea and fully developed the theory, wrote the complete manuscript, prepared figures, and reviewed the manuscript.

**Funding:** The publication of this article was funded by the Knight Foundation School of Computing and Information Sciences, Florida International University – Open Access Publishing Fund.

**Institutional Review Board Statement:** Not applicable.

**Informed Consent Statement:** Not applicable.

**Data Availability Statement:** Not applicable.

**Acknowledgments:** M.M. thanks the staff of the Knight Foundation School of Computing and Information Sciences at Florida International University for all their help and support.

**Conflicts of Interest:** The author declares no conflict of interest.

**References**

1. _. *Internet of Things (IoT): Concepts and Applications*, Alam, M., Shakil, K.A., Khan, S. (Eds.), Springer Nature Switzerland AG, 2020.
2. Greengard, S. *The Internet of Things*, The MIT Press, Boston, 2015.
3. daCosta, F. *Rethinking the Internet of Things: A Scalable Approach to Connecting Everything*, Apress Open, Santa Clara Ca, 2013.
4. _. *Building Blocks for IoT Analytics: Internet-of-Things Analytics*, Soldatos, J. (Eds.), River Publishers, Denmark, 2017.
5. Mahmuda, R.; *et al*. Profit-aware application placement for integrated Fog–Cloud computing environments, *J. Parallel Distrib. Comput*. **2020**, 135, 177–190.
6. Mohan, N.; Kangasharju, J. Edge-Fog Cloud: A Distributed Cloud for Internet of Things Computations, 2016 Cloudification of the Internet of Things (CIoT), 1-6.
7. Souza, V. B. C.; *et al*. Handling Service Allocation in Combined Fog-Cloud Scenarios, 2016 IEEE International Conference on Communications (ICC), 2016, 1-5.
8. Kallel, A., *et al*. IoT-fog-cloud based architecture for smart systems: Prototypes of autism and COVID-19 monitoring systems, Wiley, *Softw: Pract Exper*. **2021**, 51, 91–116.
9. Phillips, A. C. *Introduction to Quantum Mechanics*, Wiley, Chichester, UK, 2003.
10. _. *Applied Quantum Cryptography*, Kollmitzer, C., and Pivk, M. (Eds.) Springer-Verlag Berlin Heidelberg, 2010.
11. Cariolaro, G. *Quantum Communications: Signals and Communication Technology*, Springer, 2015.
12. Caleffi, M.; *et al*. The rise of the quantum Internet. *Computer* **2020**, 53, 06, 67–72.
13. Cacciapuoti, A. S.; *et al*. The quantum Internet: Networking challenges in distributed quantum computing. *IEEE Netw*. **2020**, 34, 1, 137–143.

14. Cacciapuoti, A. S.; *et al*. When Entanglement Meets Classical Communications: Quantum Teleportation for the Quantum Internet, *IEEE Trans. on Comm*. **2020**, 68, 6, 3808-3833.
15. Gyongyosi, L.; Imre, S. Entanglement accessibility measures for the quantum Internet. *Quant. Inf. Proc*. **2020**, 19, 115.
16. Gyongyosi, L.; Imre, S. Entanglement access control for the quantum Internet. *Quantum Inf Process*. **2019**, 18, 107.
17. Gyongyosi, L.; Imre, S. Opportunistic Entanglement Distribution for the Quantum Internet. *Sci Rep* **2019**, 9, 2219.
18. Hiskett, P. A.; *et al*. Long-distance quantum key distribution in optical fibre, *New J. Phys*. **2006**, 8, 193.
19. Ruihong, Q.; Ying, M. Research progress of quantum repeaters. IOP *J. Phys. Conf. Ser*. **2019**, 1237, 052032.
20. Polnik, M.; *el at*. Scheduling of space to ground quantum key distribution, *EPJ Quantum Technology*, **2020**, 7, 3.
21. Bedington, R.; *et al*. Progress in satellite quantum key distribution, *npj Quantum Information*, **2017**, 3, 30.
22. Ntanos, A.; *et al*. LEO Satellites Constellation-to-Ground QKD Links: Greek Quantum Communication Infrastructure Paradigm, *Photonics*, **2021**, 8, 544.
23. Liao, S-K.; *et al*. Satellite-to-ground quantum key distribution, *Nature*, **2017**, 549, 43-49.
24. Bennett, C.H., Brassard, G. Quantum cryptography: Public key distribution and coin tossing. In Proceedings of IEEE International Conference on Computers, Systems and Signal Processing, Bangalore, India, 1984, 175-179.
25. Ekert, A.K. Quantum cryptography based on Bell's theorem. *Phys. Rev. Lett.*, **1991**, 67, 6, 661–663.
26. Bennett, C.H. Quantum cryptography using any two nonorthogonal states, *Phys. Rev. Lett*., **1992**, 68, 21, 3121-3124.
27. NSA: Quantum key distribution (QKD) and quantum cryptography QC. Available online: https://www.nsa.gov/Cybersecurity/Quantum-Key-Distribution-QKD-and-Quantum-Cryptography-QC/ (accessed on 5 March 2022).
28. Sharma, V., Padakandla, A. Quantum Error Correction and Quantum Information Theory (2020). https://ece.iisc.ac.in/~spcom/2020/Slides/TutorSPCOM2020_QECC&OCryptographyOIT_Final.pdf
29. Mink, A., *et al*. Quantum Key Distribution (QKD) and Commodity Security Protocols: Introduction and Integration, *Intern. J. Network Security & Its Applications*, **2009**, 1, 2, 101-112.
30. Benenti, G, *et al*. Principles of Quantum Computation and Information I: Volume I: Basic Concepts. Singapore, World Scientific, 2008.
31. Zhou R G; Zhang X. Controlled deterministic secure semiquantum communication. *Intern. J. Theo. Phys.*, **2021**, 60, 5, 1767-1782.
32. Kravtsov K S, *et al*. Relativistic quantum key distribution system with one-way quantum communication. *Sci Rep*, **2018**, 8, 1, 1-7.
33. Bruß, D.; Lütkenhaus, N. Quantum Key Distribution: from Principles to Practicalities: A review. arxiv 1999, arXiv:9901061.
34. Makarov, V., *et al*. Effects of detector efficiency mismatch on security of quantum cryptosystems, *Phys. Rev. A*, **2006**, 74, 2, 022313.
35. Lydersen, L., *et al*. Hacking commercial quantum cryptography systems by tailored bright illumination, *Nature Photonics*, **2010**, 4, 10, 686.
36. Gerhardt, I., *et al*. Fullfield implementation of a perfect eavesdropper on a quantum cryptography system, *Nature communications*, **2011**, 2, 349.
37. Qi, B., *et al*. Time-shift attack in practical quantum cryptosystems: A review. Arxiv 2005, arXiv:0512080.
38. Weier, H., *et al*. Quantum eavesdropping without interception: an attack exploiting the dead time of single-photon detectors, *New J. Phys*, **2011**, 13, 7, 073024.
39. Bugge, A. N., *et al*. Laser damage helps the eavesdropper in quantum cryptography, *Phys. Rev. Lett*., **2014**, 112, 7, 070503.
40. Gisin, N., *et al*. Trojan-horse attacks on quantum-key-distribution systems, *Phys. Rev. A*, **2006**, 73, 2, 022320.
41. Li, H.W., *et al*. Attacking a practical quantum-key-distribution system with wavelength-dependent beam-splitter and multi-wavelength sources, *Phys. Rev. A*, **2011**, 84, 6, 062308.
42. Diamanti, E. Security and implementation of differential phase shift quantum key distribution systems. Ph.D. Thesis, CNRS 2006.
43. Holevo, A. S. Bounds for the quantity of information transmitted by a quantum communication channel, Problemy Peredachi Informatsii, **1973**, 9, 3, 3-11.
44. Fiurasek, J. Optical implementations of the optimal phase-covariant quantum cloning machine, *Phys. Rev. A*, **2003**, 67, 5, 052314.
45. Niederberger, A., *et al*. Photon-Number-Splitting versus Cloning Attacks in Practical Quantum Cryptography, Tech. report 2004.
46. Reiserer, A., *et al*. Nondestructive detection of an optical photon, *Science*, **2013**, 1246164.
47. Nogues, G., *et al*. Seeing a single photon without destroying it, *Nature*, **1999**, 400, 6741, 239.
48. Hwang, W.Y. Quantum key distribution with high loss: toward global secure communication, *Phys. Rev. Lett*., **2003**, 91, 5, 057901.
49. Lucamarini, M., *et al*. ETSI White Paper No. 27, Implementation Security of Quantum Cryptography: Introduction, challenges, solutions, 2018.
50. Tsai, C.W., *et al*. Quantum Key Distribution Networks: Challenges and Future Research Issues in Security, *Appl. Sci*. **2021**, 11, 3767.
51. Lo, H-K., *et al*. Secure quantum key distribution, *Nature Photonics*, **2014**, 8, 595-604.
52. Lu, F-Y. Practical issues of twin-field quantum key distribution, *New J. Phys.*, **2019**, 21, 123030.
53. Diamanti, E., *et al*. Practical challenges in quantum key distribution, *npj Quantum Information,* **2016**, 2, 16025.
54. Zhang, Q., *et al*. Large scale quantum key distribution: challenges and solutions, *Opt. Express*, **2018**, 26, 24260-24273.

55. Lopez-Leyva, J.A., *et al*. Quantum Cryptography in Advanced Networks, Morozov, O.G. (Eds.), Chapter: Free-Space-Optical Quantum Key Distribution Systems: Challenges and Trends, IntechOpen, 2018.
56. Fung, C-H. F., *et al*. Practical issues in quantum-key-distribution post-processing: A review. arxiv 2009, arXiv:0910.0312v2
57. Zhao, Y., *et al*. Quantum hacking: experimental demonstration of time-shift attack against practical quantum key distribution systems: A review. arxiv 2011, arXiv:0704.3253v3
58. Lovic, V. Quantum Key Distribution: Advantages, Challenges and Policy, *Cambridge Journal of Science & Policy*, **2020**, 1, 2.
59. Bacco, D., *et al*. Boosting the secret key rate in a shared quantum and classical fibre communication system. *Comm. Phys.* **2019**, 2, 140.
60. Aliro: Quantum network security: What is Quantum Key Distribution (QKD)? Available online: https://www.aliroquantum.com/blog/quantum-network-security-what-is-quantum-key-distribution-qkd?hsLang=en (accessed on 25 February 2022).
61. Tajima, A., *et al*. Quantum key distribution network for multiple applications, *Quantum Sci. Technol*. **2017**, 2, 034003.
62. Agnesi, C., *et al*. Simple quantum key distribution with qubit-based synchronization and a self-compensating polarization encoder, *Optica*, **2020**, 7, 4, 284-290.
63. Price, A.B. A quantum key distribution protocol for rapid denial of service detection, *EPJ Quantum Technol*. **2020**, 7, 8.
64. Sasaki, M. Quantum Key Distribution and Its Applications, in *IEEE Security & Privacy*, **2018**, 16, 5, 42-48.
65. Mehic, M., *et al*. Quantum Key Distribution: A Networking Perspective, *ACM Computing Surveys*, **2020**, 53, 5, 96.
66. Bennett, C. H.; *et al.* Teleporting an unknown quantum state via dual classical and Einstein–Podolsky–Rosen channels. *Phys. Rev. Lett*. **1993**, 70, 13, 1895–1899.
67. Liorni, C., *et al*. Satellite-based links for quantum key distribution: beam effects and weather dependence. *New J. Phys*. **2019**, 21, 093055.
68. Zhao, B., *et al*. Performance Analysis of Quantum Key Distribution Technology for Power Business. *Appl. Sci*. **2020**, 10, 2906.
69. Bedington, R., *et al*. Progress in satellite quantum key distribution, *npj Quantum Information*, **2017**, 3, 30.
70. Yin, J., *et al*., Entanglement-based secure quantum cryptography over 1,120 kilometres, *Nature*, **2020**, 582, 501–505.
71. Sergienko, A.V. *Quantum communications and cryptography*, CRC Taylor & Frances, Boca Raton, 2006.
72. Mastriani, M., *et al*. Bidirectional teleportation for underwater quantum communications, *Quantum Inf. Proce*, **2021**, 20, 22.
73. Mastriani, M., Iyengar, S. S. Satellite quantum repeaters for a quantum Internet. Wiley, *Quantum Engineering*, QUE55, **2020**.
74. Liu, H-Y., *et al*. Drone-based entanglement distribution towards mobile quantum networks. *Nat. Sci. Rev.*, **2020**, 7, 921–928.
75. Mastriani, M., *et al*. Satellite quantum communication protocol regardless of the weather, *Opt. and Quantum Electronics*, **2021**, 53, 181.
76. Hill, A. D., *et al*. Drone-based Quantum Key Distribution (2017). http://2017.qcrypt.net/wp-content/uploads/2017/09/Tu22.pdf
77. Sharma, S., *et al*. Tracking challenges of QKD over relay satellite, International Conference on Space Optics, Chania, Greece, Proc. of SPIE, 2018, 11180 1118062-1.
78. Zhang, Q., *et al*. Large scale quantum key distribution: challenges and solutions, *Opt. Express*, **2018**, 26, 24260-24273.
79. Vázquez-Castro, A., *et al*. Quantum Keyless Private Communication Versus Quantum Key Distribution for Space Links, *Phys. Rev. Applied*, **2021**, 16, 1, 014006.
80. Chinese 'space cleaner' spotted grabbing and throwing away old satellite. Available online: https://indianexpress.com/article/technology/science/chinese-space-cleaner-spotted-grabbing-and-throwing-away-old-satellite/ (accessed on 3 March 2022).
81. Modern spy satellites in an age of space wars (signal intelligence). Available online: https://www.dw.com/en/modern-spy-satellites-in-an-age-of-space-wars/a-54691887 (accessed on 3 March 2022).
82. Satellite cyber attack paralyzes 11GW of German wind turbines. Available online: https://www.pv-magazine.com/2022/03/01/satellite-cyber-attack-paralyzes-11gw-of-german-wind-turbines/ (accessed on 3 March 2022).
83. Space Debris and Human Spacecraft. Available online: https://www.nasa.gov/mission_pages/station/news/orbital_debris.html (accessed on 25 February 2022).
84. Orbital Objects: Learn more about satellites, space junk, and other objects floating in orbits. Available online: https://www.nationalgeographic.com/science/article/orbital-objects (accessed on 25 February 2022).
85. How space debris created the world's largest garbage dump. Available online: https://bigthink.com/hard-science/space-debris-solutions/ (accessed on 25 February 2022).
86. What is space junk and why is it a problem? Available online: https://www.nhm.ac.uk/discover/what-is-space-junk-and-why-is-it-a-problem.html (accessed on 25 February 2022).
87. About space debris. Available online: https://www.esa.int/Safety_Security/Space_Debris/About_space_debris (accessed on 25 February 2022).
88. Who owns our orbit: Just how many satellites are there in space? Available online: https://www.weforum.org/agenda/2020/10/visualizing-easrth-satellites-sapce-spacex/ (accessed on 25 February 2022).
89. SpaceX's Dark Satellites Are Still Too Bright for Astronomers. Available online: https://www.scientificamerican.com/article/spacexs-dark-satellites-are-still-too-bright-for-astronomers/ (accessed on 25 February 2022).
90. Elon Musk's SpaceX Starlink satellites disrupt almost a FIFTH of images snapped by a crucial asteroid-spotting telescope, study warns. Available online: https://www.dailymail.co.uk/sciencetech/article-10422961/SpaceX-Starlink-satellites-disrupt-FIFTH-images-telescope.html (accessed on 25 February 2022).

91. Satellite mega-constellations including SpaceX's Starlink are now a WORSE threat to astronomy than light pollution, experts warn. Available online: https://www.dailymail.co.uk/sciencetech/article-10486455/SpaceXs-Starlink-WORSE-threat-astronomy-light-pollution-experts-warn.html (accessed on 25 February 2022).
92. Satellite constellations: Astronomers warn of threat to view of Universe. Available online: https://www.bbc.com/news/science-environment-50870117 (accessed on 25 February 2022).
93. SpaceX Starlink satellites interfere with the study of dangerous asteroids. Available online: https://thetimeshub.in/spacex-starlink-satellites-interfere-with-the-study-of-dangerous-asteroids (accessed on 25 February 2022).
94. International Astronomical Union launches new center to fight satellite megaconstellation threat. Available online: https://www.space.com/iau-center-protect-astronomy-megaconstellation-threat (accessed on 10 March 2022).
95. New center to coordinate work to mitigate effect of satellite constellations on astronomy. Available online: https://spacenews.com/new-center-to-coordinate-work-to-mitigate-effect-of-satellite-constellations-on-astronomy/ (accessed on 10 March 2022).
96. Protection of the Dark and Quiet Sky from Satellite Constellation Interference. Available online: https://www.mpifr-bonn.mpg.de/announcements/2022/1 (accessed on 10 March 2022).
97. SKAO and international partners petition UN for the protection of Earth's dark and quiet skies. Available online: https://www.skatelescope.org/news/skao-and-international-partners-petition-un-for-the-protection-of-earths-dark-and-quiet-skies/
98. Orbiting robots could help fix and fuel satellites in space: Machines will soon have a go at maintaining fleet of small spacecraft orbiting Earth. Available online: https://arstechnica.com/science/2022/03/orbiting-robots-could-help-fix-and-fuel-satellites-in-space/?comments=1 (accessed on 10 March 2022).
99. Nielsen, M. A., Chuang, I. L. *Quantum Computation and Quantum Information*, Cambridge University Press, N.Y., 2004.
100. Quantis QRNG Chip. Available online: https://www.idquantique.com/random-number-generation/products/quantis-qrng-chip/ (accessed on 10 March 2022).
101. Swan, M. *Blockchain: Blueprint for a New Economy*. O'Reilly, Sebastopol, CA, 2015.
102. Vernam, G. S. Cipher Printing Telegraph Systems For Secret Wire and Radio Telegraphic Communications. *Journal of the IEEE*, **1926**, 55, 109–115.
103. Quantum Experiments at Space Scale. Available online: https://en.wikipedia.org/wiki/Quantum_Experiments_at_Space_Scale (accessed on 14 March 2022).
104. Kets. Available online: https://kets-quantum.com/quantum-key-distribution/ (accessed on 14 March 2022).
105. ISO/IEC 27001: Information Security Management. Available online: https://www.iso.org/isoiec-27001-information-security.html (accessed on 14 March 2022).
106. International Association of Classification Societies (IACS). Available online: https://iacs.org.uk/publications/recommendations/161-180/rec-166-new-corr1/ (accessed on 14 March 2022).
107. DoD Space Agency. Available online: https://spacenews.com/dod-space-agency-funds-development-of-laser-terminal-that-connects-to-multiple-satellite-at-once/ (accessed on 14 March 2022).
108. TechRepublic. Available online: https://www.techrepublic.com/article/how-an-8-character-password-could-be-cracked-in-less-than-an-hour/ (accessed on 14 March 2022).